\documentclass[12pt,graphicx]{article}

\begin{document} 

\begin{center} 
{\Large {\bf The Masses of Bosons and Usual Fermions on Supersymmetric 
$SU(3)_{C}\times SU(4)_{L}\times U(1)_{X}$ Model}}
\end{center}

\begin{center}
M. C. Rodriguez  \\
{\it Grupo de F{\'{\i}}sica Te\'{o}rica e Matem\'{a}tica F\'{\i}sica \\
Departamento de F\'{\i}sica  \\
Universidade Federal Rural do Rio de Janeiro (UFRRJ) \\
BR 465 Km 7, 23890-000 Serop\'{e}dica - RJ \\
Brasil}
\end{center}

\date{\today}

\begin{abstract}
We will study in details the masses spectrum of bosons and usual fermions  of the Minimal 
Supersymmetric $SU(3)_{C}\times SU(4)_{L}\times U(1)_{X}$ Model. 
\end{abstract}

PACS number(s): 12.60. Cn, 12.60. Jv


\section{Introduction}

Today we know, Symmetry principles have been used to both issues: classification of particles and the dynamics of the interactions. 
The gauge group $SU(N)$ has two fundamental irredutivel representation \cite{greiner,Cotaescu:1995jz,Yamatsu:2015npn}
\begin{itemize}
\item n-plete $(n)$;
\item anti-n-plete $(n^{*})$;
\end{itemize}
and its algebra can be parameterized in the following way
\begin{itemize}
\item $N-1$ diagonal generators of the Cartan algebra $D_{i}$;
\item the generators  
\end{itemize}
\begin{equation}
E^{i}_{j}= \frac{H^{i}_{j}}{\sqrt{2}}, \,\ i \neq j;
\end{equation}
related to the off-diagonal generators and they satisfy the following algera \cite{key-19,key-20}
\begin{eqnarray}
 \mbox{Tr} \left( D_{i}D_{j} \right)&=& \frac{1}{2} \delta_{ij}, \nonumber \\
 \mbox{Tr} \left( D_{i}E^{i}_{j} \right)&=&0, \nonumber \\
 \mbox{Tr} \left( E^{i}_{j}E^{k}_{l} \right)&=& \frac{1}{2} \delta^{i}_{l} \delta^{k}_{j}.   
\label{genertorssun}
\end{eqnarray}
The general method for construct model with symmetry
\begin{equation}
SU(N) \times U(1),
\end{equation}
is \cite{Cotaescu:1995jz,Palcu:2009ks,Palcu:2009ky,Palcu:2009kb}
\begin{itemize}
\item[a-)] The spinor sector must be put in pure left form;
\item[b-)] The coupling constant, $g$, is the same with the first one of the SM;
\item[c-)] At least one $Z$-like boson should satisfy the mass condition 
\end{itemize}
\begin{equation}
M_{Z}= \frac{M_{W}}{\cos\theta_{W}},
\label{condgeral}
\end{equation}
established in the SM and experimentally confirmed.

The models with $SU(4)$\footnote{$SU(4)$ is isomorphic to $SO(6)$ \cite{Yamatsu:2015npn}.} symmetry satisfy\cite{greiner,Yamatsu:2015npn}
\begin{eqnarray}
SU(4) &\subset& SU(3) \times U(1); \nonumber \\
SU(4) &\subset& SU(2) \times SU(2) \times U(1); \nonumber \\
SU(4) &\subset& SU(2) \times SU(1); \nonumber \\
\end{eqnarray}

 was built for nuclear interaction in \cite{Amati:1964np,Amati:1964tc}. Remember in the Pati, Salam, Mohapatra and Senjanovic model the strong interaction was descrebide by $SU(4)_{C}$ instead of $SU(3)_{C}$ \cite{key-4,key-5,key-6,key-7}.  Models with 
the gauge symmetry (the so called 341 models) 
\begin{equation}
SU(3)_{C} \times SU(4)_{L} \times U(1)_{X},
\label{gaugegroup}
\end{equation}
Some year later it was used to explain neutrino data \cite{su4neutrinos}. The author applied $SU(4)$ symmetry in the lepton sector to shown 
the compatibility between small neutrino mass\footnote{The neutrino mass is generated only at twoo-loop level.} \cite{su4neutrinos}
\begin{equation}
m_{\nu}\approx \frac{ff^{\prime}}{16 \pi^{2}}m_{\tau}
\ln \left( \frac{m^{2}_{\eta_{1}}}{m^{2}_{\eta_{2}}} \right) \approx 30 {\mbox eV}, 
\end{equation} 
where $(\eta_{1})$\footnote{This scalar induce $m_{\nu}$ proportional to $m_{\tau}$ and it is a $SU(2)_{L}$ singlet charged scalar.} and 
$(\eta_{2})$\footnote{It give rise to additional contribution to both $\mu_{B}$ and $m_{\nu}$ and it is a doublet of $SU(2)_{L}$.} 
arescalars with the following interactions \cite{su4neutrinos}
\begin{eqnarray}
{\cal L}^{\eta_{1}}&=&f \left( \tau \nu_{e} \right) \eta_{1}- 
f^{\prime} \left( \tau^{c}\nu^{c}_{e} \right) \eta^{*}_{1}+hc, \nonumber \\ 
{\cal L}^{\eta_{2}}&=&  f \left( \tau \nu^{c}_{e} \right) \eta_{2}+ 
f^{\prime} \left( \tau^{c}\nu_{e} \right) \eta^{*}_{2}+hc,
\end{eqnarray}
and those scalars belonging to $SU(4)$ sextet, with a hypothetical neutrino magnetic moment
\begin{equation}
\mu_{\nu} \approx (0.3-1) \times 10^{-11} \mu_{B},
\end{equation}
where 
\begin{equation}
\mu_{B}= \frac{e}{(2m_{e})},
\end{equation}
is the Bohr magneton and we would find
\begin{equation}
e \frac{m_{\nu}}{\mu_{\nu}} \approx {\cal O} \left(  m^{2}_{\eta_{1}}, m^{2}_{\eta_{2}} \right).
\end{equation}

There are interesting model given at \cite{su4b,su4b1,su4b2}
\footnote{Another interesting possibility is shown \cite{Mebarki:2019eeh}.}. The supersymmetric version of 
those models was made at \cite{susy341,Rodriguez:2007jc}. The 
best feature of this model is that it provides us with an alternative to 
the problem of the number $N_{f}$ of fermion families. These sort of models are
anomaly free only if there are equal number of quartet and anti-quartet 
(considering the color degrees of freedom), and furthermore requiring the sum of 
all fermion charges to vanish. Recently it was given bounds on dipole moments of tau-neutrino \cite{Binh:2018teg}.

\section{Non-Supersymmetric Model 341 Model without exotic charges.}
\label{sec:non-susymodel}

In the contest of the gauge symmetry defined at Eq.(\ref{gaugegroup}), the most general expression for the 
electric charge generator is a linear combination of the four diagonal
generators of the gauge group 
\begin{eqnarray}
\frac{Q}{e}&=&
\frac{1}{2} \left(  \lambda_{3}+\frac{b}{\sqrt{3}} \lambda_{8}+ \frac{c}{\sqrt{6}} \lambda_{15} \right)+ XI_{4 \times 4} 
\nonumber \\
&=&\mbox{diag} \left[ \frac{1}{2} \left( 1+ \frac{b}{3}+ \frac{c}{6}  \right)+X, 
\frac{1}{2} \left( -1+ \frac{b}{3}+ \frac{c}{6}  \right)+X, \frac{1}{2} \left( \frac{-2b}{3}+ \frac{c}{6}  \right)+X, 
\frac{-c}{4}+X \right], \nonumber \\ 
\label{chargeral}
\end{eqnarray} 
where $\lambda_{i}$, being the Gell-Mann matrices for $SU(4)$ \cite{greiner}. These matrices 
are normalized as  Tr$(\lambda_{i}\lambda_{j})=2\delta_{ij}$, $I_{4 \times 4}= \mbox{diag}(1,1,1,1)$, and  $b$ 
and $c$ are free parameters to be fixed next.

The model presented at \cite{su4b,su4b2}, we choose the following parameters
\begin{equation}
b=-1, \,\ c=-1
\end{equation}
and Eq.(\ref{chargeral}) become
\begin{eqnarray}
\frac{Q}{e}&=&\mbox{diag}(X,X-1,X,X+1).
\end{eqnarray}

We have the leptons
transforming in the lowest representation of $SU(4)_{L}$ the quartet\footnote{In the same way as proposed by 
Voloshin \cite{su4neutrinos} in order to understand the existence of neutrinos with large magnetic moment and small mass.} in the following way
\begin{equation}
L_{aL}= \left( 
\begin{array}{c}
\nu_{a}\\
l_{a}\\
\nu^{c}_{a}\\
l^{c}_{a}
\end{array}
\right)_L \sim ( {\bf 1},{\bf 4},0), \,\ a=1,2,3.
\label{quadleptons}
\end{equation}
The numbers in parenthesis refer to the $(SU(3)_{C},SU(4)_{L},U(1)_{X})$ quantum 
numbers, respectively. The sum of all leptons charges in the quartet is zero. 

We can decompose the 341 model into mutiplets of 331 model in the following way\cite{Yamatsu:2015npn,Dias:2013kma}
\begin{equation}
 L_{a_{L}}^{\prime }=\left( 
\begin{array}{c}
\nu _{a} \\ 
e_{a} \\ 
\nu _{a}^{c}
\end{array}
\right) _{L}\sim \left( \mathbf{1},\mathbf{3},-1/3\right)
\oplus e_{aL}^{c}\sim \left( \mathbf{1},\mathbf{1},+1\right)\,,
\label{declep}
\end{equation}
it leads to the SM content
\begin{equation}
l_{a_{L}}=\left( 
\begin{array}{c}
\nu _{a} \\ 
e_{a}
\end{array}
\right) _{L}\sim \left( \mathbf{1},\mathbf{2},-1 \right) \oplus \nu
_{aL}^{c}\sim \left( \mathbf{1},\mathbf{1},0 \right) ,
\end{equation}

In the quark sector, one quark family is also put in the quartet 
representation 
\begin{eqnarray}
 Q_{1L} = \left(
\begin{array}{c}
u_{1}\\
d_{1}\\
u^{\prime}\\
J
\end{array}\right)_L \sim \left( {\bf 3},{\bf 4}, \frac{2}{3} \right) , 
\label{q1l}
\end{eqnarray}
and the respective singlets are given by
\begin{eqnarray}
u^{c}_{1L} &\sim& \left({\bf3}^*,{\bf1},-\frac{2}{3}\right),\quad
d^{c}_{1L} \sim \left({\bf3}^*,{\bf1},\frac{1}{3}\right),\nonumber \\ 
u^{\prime c}_{L} &\sim& \left({\bf3}^*,{\bf1},-\frac{2}{3}\right),\quad 
J^{c}_{L} \sim \left({\bf3}^*,{\bf1},-\frac{5}{3}\right),
\label{q1r}
\end{eqnarray}
writing all the fields as left-handed; $u^{\prime}$ and $J$ are new quarks with 
charge $+2/3$ and $+5/3$ respectively \footnote{The sum of all quarks charges in the 
quartet is equal to $\frac{8}{3}$, while the sum of all quarks charges in the 
singlet is equal to $- \frac{8}{3}$, therefore the total charge 
in this first familly is equal zero as required on the introduction.}.

The others two quark generations, as we have explained in the introduction, we put in the anti-quartet 
representation \footnote{We associted three quartets to the leptons and three quartets to one family of 
quarks, now we are associating six anti-quartets, therefore we can get the cancelation the anomaly in this model.}
\begin{equation}
\begin{array}{cc}  
Q_{2 L} = \left(
\begin{array}{c}
d_{2}\\
u_{2}\\
d^{\prime}_{1}\\
j_{1}
\end{array}\right)_L \sim \left({\bf3},{\bf4}^{*},- \frac{1}{3}\right) ,\quad  
 Q_{3L} = 
      \left( \begin{array}{c} 
d_{3}\\
u_{3}\\
d^{\prime}_{2}\\
j_{2}
\end{array} \right)_{L} 
\sim \left({\bf3},{\bf4}^{*},- \frac{1}{3}\right) , 
\end{array}
\label{q23l}
\end{equation}
and also with the respective singlets,
\begin{eqnarray}
u^{c}_{\alpha L} &\sim& \left({\bf3}^*,{\bf1},-\frac{2}{3}\right),\quad
d^{c}_{\alpha L} \sim \left({\bf3}^*,{\bf1},\frac{1}{3}\right), \nonumber \\
d^{\prime c}_{\beta L} &\sim& \left({\bf3}^*,{\bf1},\frac{1}{3}\right), \quad
j^{c}_{\beta L}  \sim \left({\bf3}^*,{\bf1},\frac{4}{3}
\right) \,\ , 
\label{q23r}
\end{eqnarray}
$j_{\beta}$ and $d^{\prime}_{\beta}$, $\beta =1,2$ are new quarks with charge $-4/3$ and
$-1/3$ respectively, while $\alpha =2,3$ is the familly index for the quarks 
\footnote{The sum of all quarks charges in the quartet is equal to $- \frac{4}{3}$, 
while the sum of all quarks charges in the singlet is equal to $\frac{4}{3}$, 
therefore the total charge in the second and in the third families are equal zero.}. 
We remind that in Eqs.(\ref{q1l},\ref{q1r},\ref{q23l},\ref{q23r})
all fields are still symmetry eigenstates. The decomposition of those fields in 
terms of 331 and SM gauge group can be found at \cite{Dias:2013kma}.

On the other hand, the scalars, in quartet, which are necessary to generate the 
quark masses are
\begin{eqnarray} 
\eta&=&\left(
\begin{array}{c}
\eta_{1}^{0} \\ \eta^{-}_{1} \\ \eta^{0}_{2} \\ \eta^{+}_{2}
\end{array}
\right) \sim \left( {\bf 1},{\bf 4},0 \right), \,\
\rho=\left(
\begin{array}{c}
\rho_{1}^{+} \\ \rho^{0} \\ \rho^{+}_{2} \\ \rho^{++}
\end{array}
\right) \sim \left( {\bf 1},{\bf 4},1 \right), \nonumber \\
\chi&=&\left(
\begin{array}{c}
\chi_{1}^{-} \\ \chi^{--} \\ \chi^{-}_{2} \\ \chi^{0}
\end{array}
\right) \sim \left( {\bf 1},{\bf 4}, -1 \right).
\label{4t} 
\end{eqnarray}
The Yukawa couplings for the quark sector are
\begin{eqnarray}
- {\cal L}^{\eta \,\ \rho \,\ \chi}_{Yukawa}  & = & 
h^{t}\, \overline{Q_{3 L} }\,  \eta u_{3 R}   + h^{b}\, \overline{Q_{3 L} } \rho d_{3 R}   +
 h^{J}\, \overline{Q_{3 L} }\,  \chi\, J_{ R}    \nonumber \\
& + &  h^{d2}_{\alpha \beta}\overline{Q_{\alpha  L} } \eta^\dag d_{\beta R}   +
h^{u2}_{\alpha \beta}\overline{Q_{\alpha  L} } \rho^\dag u_{\beta R}
+ h^{j}_{\alpha \beta}\overline{Q_{\alpha  L} } \chi^\dag j_{\beta R}
\nonumber \\
&+& hc.
\label{m4eq468a}
\end{eqnarray}

In order to avoid mixing among primed and unprimed
quarks, we have to introduce an extra scalar transforming like $\eta$ but with different 
vacuum expectation value (VEV)
\begin{equation}
\phi=\left(
\begin{array}{c}
\phi_{1}^{0} \\ \phi^{-}_{1} \\ \phi^{0}_{2} \\ \phi^{+}_{2}
\end{array}
\right) \sim \left( {\bf 1},{\bf 4},0 \right) \,\ .
\label{4tf}
\end{equation}
it generate the following coupling
\begin{eqnarray}
- {\cal L}^{\phi}_{Yukawa}  & = & 
h^{u^{\prime}}\overline{Q_{3 L} }\,  \phi \,  u^{\prime}_{R}   +
 h^{d^{\prime}}_{\alpha \beta}\overline{Q_{\alpha  L} } \phi^\dag d^{\prime}_{\beta R}   +  hc.
\label{m4eq468c}
\end{eqnarray}

The vev of our scalars are given by
\begin{eqnarray}
\langle\eta\rangle^{T}&=& \left( \frac{v}{\sqrt{2}},0,0,0 \right)^{T}, \,\ 
\langle\rho\rangle^{T}= \left( 0, \frac{u}{\sqrt{2}},0,0 \right)^{T}, \nonumber \\
\langle\phi\rangle^{T}&=& \left( 0,0,\frac{z}{\sqrt{2}},0 \right)^{T}, \,\ 
\langle\chi\rangle^{T}= \left( 0,0,0, \frac{w}{\sqrt{2}} \right)^{T}. \,\  \nonumber \\
\label{vev341}
\end{eqnarray}
The Yukawa couplings for the quark sector, see Eq.(\ref{m4eq468a},\ref{m4eq468c}),  are
\begin{eqnarray}
- {\cal L}^{q}_{Yukawa}  & = & {\cal L}^{\eta \,\ \rho \,\ \chi}_{Yukawa}+ {\cal L}^{\phi}_{Yukawa},
\label{m4eq468}
\end{eqnarray}
then the quarks get masses as follows \cite{Long:2016lmj}
\begin{eqnarray} 
m_{u_{3}} & = & h^t \frac{u}{\sqrt{2}}, \,  m_{d_{3}} = h^b \frac{v}{\sqrt{2}} \, ,
m_{u^{\prime}} = h^{u^{\prime}} \frac{\omega}{\sqrt{2}}\, , m_{J} = h^{J} \frac{V}{\sqrt{2}}\, ,\nonumber \\
(m_{d_2})_{ \alpha \beta } & = & h^{d2}_{\alpha \beta} \frac{u}{\sqrt{2}}\, ,  
(m_{u_2})_{ \alpha \beta } = -h^{u2}_{\alpha \beta} \frac{v}{\sqrt{2}}\, ,
 (m_{d^{\prime}})_{ \alpha \beta} = h^{d^{\prime}}_{\alpha \beta} \frac{\omega}{\sqrt{2}}\, , 
 (m_{D^ \prime_2})_{ \alpha \beta } =  h^{j}_{\alpha \beta}\ \frac{V}{\sqrt{2}}\, . \nonumber \\
\label{m4eq506}
\end{eqnarray}

However  we know 
\begin{equation}
\overline{L^{c}}L= 4 \times 4= 6_{A} \oplus 10_{S},
\end{equation}
however $6_{A}$ will leave some leptons massless and therefore for get mass for all the charged leptons we need to introduce the following 
symmetric anti-decuplet \cite{su4neutrinos,su4b,su4b1}
\begin{equation}
H=\left(
\begin{array}{cccc}
H^{0}_{1}\, &\, H^{+}_{1}\, &\, H_{2}^{0}\, &\, H_{2}^{-} \\
H^{+}_{1}\, &\, H_{1}^{++}\, &\, H_{3}^{+}\, &\, H_{3}^{0} \\
H^{0}_{2}\, &\, H^{+}_{3}\, &\, H^{0}_{4}\, &\, H^{-}_{4} \\
H^{-}_{2}\, &\, H^{0}_{3}\, &\, H^{-}_{4}\, &\, H^{--}_{2}
\end{array}
\right) \sim \left( {\bf 1},{\bf 10}^{*},0 \right).
\label{sextet}
\end{equation}
The Yukawa interaction for the lepton  is given by \cite{Long:2016lmj}
\begin{eqnarray} 
- {\cal L}_{Yukawa}^{l} & = & h^l_{a b}  \overline{f_{a L} } H  f^c_{b R}  +hc,
\label{m4eq4358}
\end{eqnarray}
then
 \begin{equation}  
\langle H \rangle  =\frac{1}{2}\left(%
\begin{array}{cccc}
0 & 0 &  y & 0 \\
0 & 0   & 0 &  x \\
y & 0  & 0 & 0 \\
0 & x  & 0 & 0
\end{array}\,
\right)\, . \label{Higg14}
\end{equation}
The charged leptons get mass matrix  given by
\begin{equation} 
(m_{l})_ {a b} =  \frac{h^{l}_{a b}}{\sqrt{2}} \langle H_{3}^{0} \rangle = \frac{h^{l}_{a b} \, x}{2}
   \, .\label{m4eq448}
\end{equation}
The neutrinos obtain  the  Dirac mass  by $\langle H_{2}^{0} \rangle$  in the same Yukawa coupling  matrix:
\begin{equation} 
(m_{\nu})_{ a b} =   \frac{h^{l}_{a b}}{\sqrt{2}} \langle H_{2}^{0} \rangle = \frac{h^{l}_{a b}\, y}{2}.
 \label{m4eq448a}
\end{equation}
The neutrino Majorana mass will follow from $\langle H_{1}^{0} \rangle$ and $\langle H_{4}^{0} \rangle$.

The electroweak gauge bosons of this theory are defined at Eq.(\ref{defcharggaugebosons}) and they get the following masses, see 
our Eq.(\ref{gaugebosonsu(4)quad}) \cite{su4b,su4b1} 
\begin{eqnarray}
M^{2}_W&=&\frac{g^{2}}{4} \left( v^{2}+u^{2}+2x^{2} \right), \quad
M^{2}_{V_{1}}=\frac{g^{2}}{4} \left( z^{2}+u^{2}+2x^{2} \right), \nonumber \\
M^{2}_{V_{2}}&=&\frac{g^{2}}{4} \left( v^{2}+w^{2}+2x^{2} \right),\quad
M^{2}_{V_{3}}=\frac{g^{2}}{4} \left( z^{2}+w^{2}+2x^{2} \right), \nonumber \\
M^{2}_{X}&=& \frac{g^{2}}{8} \left( v^{2}+x^{2} \right),  \quad
M^{2}_{U}= \frac{g^{2}}{4} \left( v^{2}+w^{2}+4x^{2} \right).
\label{wmass}
\end{eqnarray}

The mass matrix for the neutral vector bosons (up to a factor
$g^{2}/4$) in the $\left( V^{3},V^{8},V^{15},V^{\prime} \right)$ basis is \cite{su4b,su4b1}
\begin{equation}
\left(
\begin{array}{llll}
v^2\!+\!u^2\!+\!v''^2\, &\, \frac{1}{\sqrt3}(v^2\!-\!u^2\!-\!v''^2)\, &\,
\frac{1}{\sqrt6}(v^2\!-\!u^2\!+\!2v''^2)\, &\, -2tu^2 \\
 \frac{1}{\sqrt3}(v^2\!-\!u^2\!-\!v''^2)\, &\,
\frac{1}{3}(v^2\!+\!4v'^2\!+\!u^2\!+\!v''^2)\, &
\frac{1}{3\sqrt2}(v^2\!-\!2v'^2\!+\!u^2\!-\!2v''^2)\, &\,
\frac{2}{\sqrt3}tu^2 \\
\frac{1}{\sqrt6}(v^2\!-\!u^2\!+\!2v''^2)\, &\,
\frac{1}{3\sqrt2}(v^2\!-\!2v'^2\!+\!u^2\!-\!2v''^2)
&\, \frac{1}{6}(v^2+v'^2+u^2+9w^2+4v''^2)\, &\,
\frac{2}{\sqrt6}t(u^2+3w^2) \\
-2tu^2\, &\, \frac{2}{\sqrt3}tu^2\, &\, \frac{2}{\sqrt6}t(u^2\!+\!3w^2)\,
&4t^2(u^2\!+\!w^2)
\end{array}
\right)
\label{zmass}
\end{equation}
where 
\begin{equation}
t\equiv \frac{g^{\prime}}{g}.
\end{equation} 

The matrix in (\ref{zmass}) has determinant
equal to zero as it must be
in order to have a massless photon. There are four neutral bosons: a
massless $\gamma$ and three massive
ones: $Z,Z^{\prime},Z^{\prime \prime}$ such that 
$M_Z<M_{Z^{\prime}}<M_{Z^{\prime \prime}}$. The lightest one,
say $Z$, corresponds to the neutral boson of the Standard Model.

\section{Supersymmetric 341 Model}

The supersymmetric version of the model presented above was made at \cite{susy341,Rodriguez:2007jc}, We will introduce the following chiral superfields associated with leptons, see Eq.(\ref{quadleptons}), 
quarks, see Eqs.(\ref{q1l},\ref{q1r},\ref{q23l},\ref{q23r}), and 
usual scalars, see Eqs.(\ref{4t},\ref{4tf}), of each chiral superfield and anti-chiral supermultiplet 
is presented in the Tabs.(\ref{lfermionnmssm},\ref{rfermionnmssm}), 
respectively. 

\begin{table}[h]
\begin{center}
\begin{tabular}{|c|c|c|}
\hline 
$\mbox{ Chiral Superfield} $ & $\mbox{ Fermion} $ & $\mbox{ Scalar} $ \\
\hline
$\hat{L}_{iL}=( \hat{\nu}_{i}, \hat{l}_{i}, \hat{\nu}^{c}_{i}, \hat{l}^{c}_{i})^{T}_{L}
\sim({\bf 1},{\bf4},0)$ & 
$L_{iL}=(\nu_{i},l_{i}, \nu^{c}_{i},l^{c}_{i})^{T}_{L}$ & 
$\tilde{L}_{iL}=( \tilde{\nu}_{i}, \tilde{l}_{i}, \tilde{\nu}^{c}_{i}, \tilde{l}^{c}_{i})^{T}_{L}$ \\
\hline
$\hat{Q}_{3L}= \left( \hat{u}_{3}, \hat{d}_{3}, \hat{u}^{\prime}, \hat{J})^{T}_{L}\sim
({\bf 3},{\bf4},+ \left( \frac{2}{3} \right) \right)$ & 
$Q_{3L}=(u_{3},d_{3},u^{\prime},J)^{T}_{L}$ & 
$\tilde{Q}_{3L}=(\tilde{u}_{3}, \tilde{d}_{3}, \tilde{u}^{\prime}, \tilde{J})^{T}_{L}$ \\ 
\hline
$\hat{\eta}=( \hat{\eta}^{0}_{1}, \hat{\eta}^{-}_{1}, \hat{\eta}^{0}_{2}, \hat{\eta}^{+}_{2})^{T} \sim({\bf 1},{\bf4},0)$ & 
$\eta=( \eta^{0}_{1}, \eta^{-}_{1}, \eta^{0}_{2}, \eta^{+}_{2})^{T}$ & 
$\tilde{\eta}=( \tilde{\eta}^{0}_{1}, \tilde{\eta}^{-}_{1}, \tilde{\eta}^{0}_{2}, \tilde{\eta}^{+}_{2})^{T}$ \\
\hline
$\hat{\rho}=( \hat{\rho}^{+}_{1}, \hat{\rho}^{0}, \hat{\rho}^{+}_{2}, \hat{\rho}^{++})^{T} \sim({\bf 1},{\bf4},+1)$ & 
$\rho=( \rho^{+}_{1}, \rho^{0}, \rho^{+}_{2}, \rho^{++})^{T}$ & 
$\tilde{\rho}=( \tilde{\rho}^{+}_{1}, \tilde{\rho}^{0}, \tilde{\rho}^{+}_{2}, \tilde{\rho}^{++})^{T}$ \\
\hline
$\hat{\chi}=( \hat{\chi}^{-}_{1}, \hat{\chi}^{--}, \hat{\chi}^{-}_{2}, \hat{\chi}^{0})^{T} \sim({\bf 1},{\bf4},-1)$ & 
$\chi=( \chi^{-}_{1}, \chi^{--}, \chi^{-}_{2}, \chi^{0})^{T}$ & 
$\tilde{\chi}=( \tilde{\chi}^{-}_{1}, \tilde{\chi}^{--}, \tilde{\chi}^{-}_{2}, \tilde{\chi}^{0})^{T}$ \\
\hline
$\hat{\phi}=( \hat{\phi}^{0}_{1}, \hat{\phi}^{-}_{1}, \hat{\phi}^{0}_{2}, \hat{\phi}^{+}_{2})^{T} \sim({\bf 1},{\bf4},0)$ & 
$\phi=( \phi^{0}_{1}, \phi^{-}_{1}, \phi^{0}_{2}, \phi^{+}_{2})^{T}$ & 
$\tilde{\phi}=( \tilde{\phi}^{0}_{1}, \tilde{\phi}^{-}_{1}, \tilde{\phi}^{0}_{2}, \tilde{\phi}^{+}_{2})^{T}$ \\
\hline
$\hat{H}^{\prime}=\left(
\begin{array}{cccc}
\hat{H}^{\prime 0}_{1}\, &\, \hat{H}^{\prime -}_{1}\, &\, 
\hat{H}_{2}^{\prime 0}\, &\, \hat{H}_{2}^{\prime +} \\
\hat{H}^{\prime -}_{1}\, &\, \hat{H}_{1}^{\prime --}\, &\, 
\hat{H}_{3}^{\prime -}\, &\, \hat{H}_{3}^{\prime 0} \\
\hat{H}^{\prime 0}_{2}\, &\, \hat{H}^{\prime -}_{3}\, &\, 
\hat{H}^{\prime 0}_{4}\, &\, \hat{H}^{\prime +}_{4} \\
\hat{H}^{\prime +}_{2}\, &\, \hat{H}^{\prime 0}_{3}\, &\, 
\hat{H}^{\prime +}_{4}\, &\, \hat{H}^{\prime ++}_{2}
\end{array}
\right) \sim \left( {\bf 1},{\bf 10},0 \right)$ &
$\tilde{H}^{\prime}=\left(
\begin{array}{cccc}
\tilde{H}^{\prime 0}_{1}\, &\, \tilde{H}^{\prime -}_{1}\, &\, 
\tilde{H}_{2}^{\prime 0}\, &\, \tilde{H}_{2}^{\prime +} \\
\tilde{H}^{\prime -}_{1}\, &\, \tilde{H}_{1}^{\prime --}\, &\, 
\tilde{H}_{3}^{\prime -}\, &\, \tilde{H}_{3}^{\prime 0} \\
\tilde{H}^{\prime 0}_{2}\, &\, \tilde{H}^{\prime -}_{3}\, &\, 
\tilde{H}^{\prime 0}_{4}\, &\, \tilde{H}^{\prime +}_{4} \\
\tilde{H}^{\prime +}_{2}\, &\, \tilde{H}^{\prime 0}_{3}\, &\, 
\tilde{H}^{\prime +}_{4}\, &\, \tilde{H}^{\prime ++}_{2}
\end{array}
\right)$ & 
$H^{\prime}=\left(
\begin{array}{cccc}
H^{\prime 0}_{1}\, &\, H^{\prime -}_{1}\, &\, H_{2}^{\prime 0}\, &\, 
H_{2}^{\prime +} \\
H^{\prime -}_{1}\, &\, H_{1}^{\prime --}\, &\, H_{3}^{\prime -}\, &\, 
H_{3}^{\prime 0} \\
H^{\prime 0}_{2}\, &\, H^{\prime -}_{3}\, &\, H^{\prime 0}_{4}\, &\, 
H^{\prime +}_{4} \\
H^{\prime +}_{2}\, &\, H^{\prime 0}_{3}\, &\, H^{\prime +}_{4}\, &\, 
H^{\prime ++}_{2}
\end{array}
\right)$ \\
\hline
\end{tabular}
\end{center}
\caption{\small Particle content in the chiral superfields in MSUSY341 
and we neglected the color indices and $i=1,2,3$. In parenthesis it appears 
the transformations properties under the respective factors 
$(SU(3)_{C},SU(4)_{L},U(1)_{X})$.}
\label{lfermionnmssm}
\end{table}

\begin{table}[h]
\begin{center}
\begin{tabular}{|c|c|c|}
\hline 
$\mbox{ Anti-Chiral Superfield} $ & $\mbox{ Fermion} $ & $\mbox{ Scalar} $ \\
\hline
$\hat{Q}_{\alpha L}=(\hat{d}_{\alpha}, \hat{u}_{\alpha}, \hat{d}^{\prime}_{\alpha}, \hat{j}_{\alpha})^{T}_{L}
\sim \left( {\bf 3},{\bf4^{*}},- \left( \frac{1}{3} \right) \right)$ & 
$Q_{\alpha L}=(d_{\alpha},u_{\alpha},d^{\prime}_{\alpha},j_{\alpha})^{T}_{L}$ & 
$\tilde{Q}_{\alpha L}=
(\tilde{d}_{\alpha}, \tilde{u}_{\alpha}, \tilde{d}^{\prime}_{\alpha}, \tilde{j}_{\alpha})^{T}_{L}$ \\
\hline
$\hat{u}^{c}_{iL}\sim \left( {\bf 3^{*}},{\bf1},
- \left( \frac{2}{3} \right) \right)$ & $u^{c}_{iL}\equiv \bar{u}_{iR}$ & 
$\tilde{u}^{c}_{iL}$ \\ 
\hline
$\hat{d}^{c}_{iL}\sim \left( {\bf \bar{3}},{\bf1},
+ \left( \frac{1}{3} \right) \right)$ & 
$d^{c}_{iL}\equiv \bar{d}_{iR}$ & 
$\tilde{d}^{c}_{iL}$ \\ 
\hline
$\hat{J}^{c}_{L}\sim \left( {\bf \bar{3}},{\bf1},
- \left( \frac{5}{3} \right) \right)$ & 
$J^{c}_{L}\equiv \bar{J}_{R}$ & 
$\tilde{J}^{c}_{L}$   \\
\hline
$\hat{j}^{c}_{\alpha L}\sim \left( {\bf \bar{3}},{\bf1},
+ \left( \frac{4}{3} \right) \right)$ & 
$j^{c}_{\alpha L}\equiv \bar{j}_{\alpha R}$ & 
$\tilde{j}^{c}_{\alpha L}$   \\
\hline
$\hat{u}^{\prime}_{L}\sim \left( {\bf 3^{*}},{\bf1},
- \left(- \frac{2}{3} \right) \right)$ & $u^{\prime}_{L}\equiv \bar{u}^{\prime}_{R}$ & 
$\tilde{u}^{\prime}_{L}$ \\
\hline
$\hat{d}^{\prime}_{\alpha L}\sim \left( {\bf 3^{*}},{\bf1},
- \left( \frac{1}{3} \right) \right)$ & $d^{\prime}_{\alpha L}\equiv \bar{d}^{\prime}_{\alpha R}$ & 
$\tilde{d}^{\prime}_{\alpha L}$ \\
\hline
$\hat{\eta}^{\prime}=( \hat{\eta}^{\prime 0}_{1}, \hat{\eta}^{\prime +}_{1}, \hat{\eta}^{\prime 0}_{2}, \hat{\eta}^{\prime -}_{2})^{T} 
\sim({\bf 1},{\bf4^{*}},0)$ & 
$\eta^{\prime}=( \eta^{\prime 0}_{1}, \eta^{\prime +}_{1}, \eta^{\prime 0}_{2}, \eta^{\prime -}_{2})^{T}$ & 
$\tilde{\eta}^{\prime}=( \tilde{\eta}^{\prime 0}_{1}, \tilde{\eta}^{\prime +}_{1}, \tilde{\eta}^{\prime 0}_{2}, \tilde{\eta}^{\prime -}_{2})^{T}$ \\
\hline
$\hat{\rho}^{\prime}=( \hat{\rho}^{\prime -}_{1}, \hat{\rho}^{\prime 0}, \hat{\rho}^{\prime -}_{2}, \hat{\rho}^{\prime --})^{T} \sim({\bf 1},{\bf4^{*}},-1)$ & 
$\rho^{\prime}=( \rho^{\prime -}_{1}, \rho^{\prime 0}, \rho^{\prime -}_{2}, \rho^{\prime --})^{T}$ & 
$\tilde{\rho}^{\prime}=( \tilde{\rho}^{\prime -}_{1}, \tilde{\rho}^{\prime 0}, \tilde{\rho}^{\prime -}_{2}, \tilde{\rho}^{\prime --})^{T}$ \\
\hline
$\hat{\chi}^{\prime}=( \hat{\chi}^{\prime +}_{1}, \hat{\chi}^{\prime ++}, \hat{\chi}^{\prime +}_{2}, \hat{\chi}^{\prime 0})^{T} \sim({\bf 1},{\bf4^{*}},+1)$ & 
$\chi^{\prime}=( \chi^{\prime +}_{1}, \chi^{\prime ++}, \chi^{\prime +}_{2}, \chi^{\prime 0})^{T}$ & 
$\tilde{\chi}^{\prime}=( \tilde{\chi}^{\prime +}_{1}, \tilde{\chi}^{\prime ++}, \tilde{\chi}^{\prime +}_{2}, \tilde{\chi}^{\prime 0})^{T}$ \\
\hline
$\hat{\phi}^{\prime}=( \hat{\phi}^{\prime 0}_{1}, \hat{\phi}^{\prime +}_{1}, \hat{\phi}^{\prime 0}_{2}, \hat{\phi}^{\prime -}_{2})^{T} 
\sim({\bf 1},{\bf4^{*}},0)$ & 
$\phi^{\prime}=( \phi^{\prime 0}_{1}, \phi^{\prime +}_{1}, \phi^{\prime 0}_{2}, \phi^{\prime -}_{2})^{T}$ & 
$\tilde{\phi}^{\prime}=( \tilde{\phi}^{\prime 0}_{1}, \tilde{\phi}^{\prime +}_{1}, \tilde{\phi}^{\prime 0}_{2}, \tilde{\phi}^{\prime -}_{2})^{T}$ \\
\hline
$\hat{H}=\left(
\begin{array}{cccc}
\hat{H}^{0}_{1}\, &\, \hat{H}^{+}_{1}\, &\, \hat{H}_{2}^{0}\, &\, 
\hat{H}_{2}^{-} \\
\hat{H}^{+}_{1}\, &\, \hat{H}_{1}^{++}\, &\, \hat{H}_{3}^{+}\, &\, 
\hat{H}_{3}^{0} \\
\hat{H}^{0}_{2}\, &\, \hat{H}^{+}_{3}\, &\, \hat{H}^{0}_{4}\, &\, 
\hat{H}^{-}_{4} \\
\hat{H}^{-}_{2}\, &\, \hat{H}^{0}_{3}\, &\, \hat{H}^{-}_{4}\, &\, 
\hat{H}^{--}_{2}
\end{array}
\right)  \sim \left( {\bf 1},{\bf 10^{*}},0 \right)$ &
$\tilde{H}=\left(
\begin{array}{cccc}
\tilde{H}^{0}_{1}\, &\, \tilde{H}^{+}_{1}\, &\, \tilde{H}_{2}^{0}\, &\, 
\tilde{H}_{2}^{-} \\
\tilde{H}^{+}_{1}\, &\, \tilde{H}_{1}^{++}\, &\, \tilde{H}_{3}^{+}\, &\, 
\tilde{H}_{3}^{0} \\
\tilde{H}^{0}_{2}\, &\, \tilde{H}^{+}_{3}\, &\, \tilde{H}^{0}_{4}\, &\, 
\tilde{H}^{-}_{4} \\
\tilde{H}^{-}_{2}\, &\, \tilde{H}^{0}_{3}\, &\, \tilde{H}^{-}_{4}\, &\, 
\tilde{H}^{--}_{2}
\end{array}
\right) $ & 
$H=\left(
\begin{array}{cccc}
H^{0}_{1}\, &\, H^{+}_{1}\, &\, H_{2}^{0}\, &\, H_{2}^{-} \\
H^{+}_{1}\, &\, H_{1}^{++}\, &\, H_{3}^{+}\, &\, H_{3}^{0} \\
H^{0}_{2}\, &\, H^{+}_{3}\, &\, H^{0}_{4}\, &\, H^{-}_{4} \\
H^{-}_{2}\, &\, H^{0}_{3}\, &\, H^{-}_{4}\, &\, H^{--}_{2}
\end{array}
\right) $ \\
\hline
\end{tabular}
\end{center}
\caption{\small Particle content in the anti-chiral superfields in 
MSUSY341 and $\alpha =1,2$ and $i=1,2,3$.}
\label{rfermionnmssm}
\end{table} 

The vev, see our Eq.(\ref{vev341}), of our scalars are given by
\begin{eqnarray}
\langle\eta\rangle^{T}&=& \left( \frac{v}{\sqrt{2}},0,0,0 \right)^{T}, \,\ 
\langle\rho\rangle^{T}= \left( 0, \frac{u}{\sqrt{2}},0,0 \right)^{T}, \,\
\langle\phi\rangle^{T}= \left( 0,0,\frac{z}{\sqrt{2}},0 \right)^{T}, \nonumber \\ 
\langle\chi\rangle^{T}&=& \left( 0,0,0, \frac{w}{\sqrt{2}} \right)^{T}, \,\ 
\langle H \rangle  =\frac{1}{2}\left(%
\begin{array}{cccc}
0 & 0 &  y & 0 \\
0 & 0   & 0 &  x \\
y & 0  & 0 & 0 \\
0 & x  & 0 & 0
\end{array}\,
\right), \nonumber \\
\langle\eta^{\prime}\rangle^{T}&=& \left( \frac{v^{\prime}}{2},0,0,0 \right)^{T}, \,\ 
\langle\rho^{\prime}\rangle^{T}= \left( 0, \frac{u^{\prime}}{\sqrt{2}},0,0 \right)^{T}, \,\
\langle\phi^{\prime}\rangle^{T}= \left( 0,0,\frac{z^{\prime}}{\sqrt{2}},0 \right)^{T}, \nonumber \\
\langle\chi\rangle^{T}&=& \left( 0,0,0, \frac{w^{\prime}}{\sqrt{2}} \right)^{T}, \,\
\langle H_3^{\prime 0}\rangle= \frac{x^{\prime}}{\sqrt{2}}, \,\ 
\langle H \rangle  =\frac{1}{2}\left(%
\begin{array}{cccc}
0 & 0 &  y^{\prime} & 0 \\
0 & 0   & 0 &  x^{\prime} \\
y^{\prime} & 0  & 0 & 0 \\
0 & x^{\prime}  & 0 & 0
\end{array}\,
\right). \,\  \nonumber \\
\label{vev}
\end{eqnarray}

Concerning the gauge
bosons and their superpartners, if we denote the gluons by $g^b$ the respective
superparticles, the gluinos, are denoted by $\lambda^b_{C}$, with 
$b=1, \ldots,8$; and in the electroweak sector we have
$V^{a}$, with $a=1, \ldots, 15$; the gauge boson of $SU(4)_{L}$, and their gauginos partners  
$\lambda^{a}_{A}$; finally we have the gauge boson of 
$U(1)_{N}$, denoted by $V^\prime$, and its supersymmetric partner $\lambda_{B}$.

\begin{table}[h]
\begin{center}
\begin{tabular}{|c|c|c|c|c|}
\hline 
${\rm{Vector \,\ Superfield}}$ & ${\rm{Gauge \,\ Bosons}}$ & ${\rm{Gaugino}}$ & $\mbox{D field}$ & 
${\rm Gauge \,\ constant}$ \\
\hline 
$\hat{V}^{a}_{C}$ & $g^{a}$ & $\tilde{g}^{a}$ & $D^{a}_{C}$ & $g_{s}$ \\
\hline 
$\hat{V}^{i}$ & $V^{i}$ & $\tilde{V}^{i}$ & $D^{i}$ & $g$ \\
\hline
$\hat{V}^{\prime}$ & $V^{\prime}$ & $\tilde{V}^{\prime}$ & $D$ & $g^{\prime}$ \\
\hline
\end{tabular}
\end{center}
\caption{\small Particle content in the vector superfields in any 
Supersymmetric model with 3-4-1 gauge symmetry.}
\label{gaugemsusy341}
\end{table}

In this model we can choose the following $R$-charges \cite{susy341,Rodriguez:2007jc}
\begin{eqnarray}
n_{\eta^{\prime}}&=&n_{\phi}=n_{\rho}=n_{H^{\prime}}=1, \,\ 
n_{\eta}=n_{\phi}=n_{\rho^{\prime}}=n_{H}=-1, \nonumber \\
n_{L}&=&n_{Q_{1}}=n_{Q_{\alpha}}=n_{d_{i}}=n_{u^{\prime}}= \frac{1}{2}, \, 
n_{J}=n_{j}=- \frac{1}{2}, \nonumber \\
n_{u_{i}}&=&n_{d^{\prime}}=- \frac{3}{2}, \,\ n_{\chi}=n_{\chi^{\prime}}=0,
\label{rdiscsusy331rn} 
\end{eqnarray}
with this $R$-charge assignment, we get that all the usual particle in the 
341 model has $R$-charge equal one while their superpartner has $R$-charge 
opposite, as happen in the MSSM.

The terms in the superpotential that satisfy the $R$-parity, $W_{2RC}+W_{3RC}$, are given by the following terms
\begin{eqnarray}
W_{2RC}&=&\mu_{ \eta} \hat{ \eta} \hat{ \eta}^{\prime}+
\mu_{ \phi} \hat{ \phi} \hat{ \phi}^{\prime}+
\mu_{ \rho} \hat{ \rho} \hat{ \rho}^{\prime}+ 
\mu_{ \chi} \hat{ \chi} \hat{ \chi}^{\prime}+
\mu_{H} \hat{H} \hat{H}^{\prime}, \nonumber \\
W_{3RC}&=& \lambda_{2ab} \epsilon \hat{L}_{aL} \hat{L}_{bL} \hat{ \eta}+
\lambda_{4ab} \hat{L}_{aL} \hat{L}_{bL} \hat{H}+ 
f_{1} \epsilon \hat{ \rho} \hat{ \chi} \hat{ \eta}+
f_{6} \hat{ \chi} \hat{ \rho} \hat{H}+ 
f^{\prime}_{1}\epsilon \hat{\rho}^{\prime}\hat{\chi}^{\prime}\hat{\eta}^{\prime} \nonumber \\ &+&
f^{\prime}_{6}\hat{\chi}^{\prime}\hat{\rho}^{\prime}\hat{H}^{\prime}+  
\kappa_{1i} \hat{Q}_{3L} \hat{\eta}^{\prime} \hat{u}^{c}_{iL}+
\kappa^{\prime}_{2} \hat{Q}_{3L} \hat{\phi}^{\prime} \hat{u}^{\prime c}_{L}+
\kappa_{3i} \hat{Q}_{3L} \hat{\rho}^{\prime} \hat{d}^{c}_{iL}+
\kappa_{4} \hat{Q}_{3L} \hat{\chi}^{\prime} \hat{J}^{c}_{L} \nonumber \\ &+&
\kappa_{5 \alpha i} \hat{Q}_{\alpha L} \hat{\rho} \hat{u}^{c}_{iL}+
\kappa_{6 \alpha i} \hat{Q}_{\alpha L} \hat{\eta} \hat{d}^{c}_{iL}+
\kappa^{\prime}_{7 \alpha \beta} \hat{Q}_{\alpha L} \hat{\phi} \hat{d}^{\prime c}_{ \beta L} +
\kappa_{8 \alpha \beta} \hat{Q}_{\alpha L} \hat{\chi} \hat{j}^{c}_{\beta L}.
\label{sprc}
\end{eqnarray}

While the $R$-parity violating terms are given by $W_{2RV}+W_{3RV}$, where
\begin{eqnarray}
W_{2RV}&=&\mu_{0a}\hat{L}_{aL} \hat{ \eta}^{\prime}+
\mu_{1a}\hat{L}_{aL} \hat{ \phi}^{\prime}+
\mu_{2} \hat{ \eta} \hat{ \phi}^{\prime}+
\mu_{3} \hat{ \phi} \hat{ \eta}^{\prime}, \nonumber \\
W_{3RV}&=& \lambda_{1abc} \epsilon \hat{L}_{aL} \hat{L}_{bL} \hat{L}_{cL}+
\lambda_{3ab} \epsilon \hat{L}_{aL} \hat{L}_{bL} \hat{ \phi}+
\lambda_{5a} \epsilon \hat{L}_{aL} \hat{\chi} \hat{\rho}+
f_{2} \epsilon \hat{ \rho} \hat{ \chi} \hat{ \phi} \nonumber \\ &+&
f_{3} \hat{ \eta} \hat{ \eta} \hat{H}+
f_{4} \hat{ \eta} \hat{ \phi} \hat{H}+
f_{5} \hat{ \phi} \hat{ \phi} \hat{H}+
f^{\prime}_{2}\epsilon \hat{\rho}^{\prime}\hat{\chi}^{\prime}\hat{\phi}^{\prime}+
f^{\prime}_{3}\hat{\eta}^{\prime}\hat{\eta}^{\prime}\hat{H}^{\prime}+
f^{\prime}_{4}\hat{\eta}^{\prime}\hat{\phi}^{\prime}\hat{H}^{\prime}+
f^{\prime}_{5}\hat{\phi}^{\prime}\hat{\phi}^{\prime}\hat{H}^{\prime} \nonumber \\ &+& 
\kappa^{\prime}_{1} \hat{Q}_{3L} \hat{\eta}^{\prime} \hat{u}^{\prime c}_{L}+
\kappa_{2i} \hat{Q}_{3L} \hat{\phi}^{\prime} \hat{u}^{c}_{iL}+
\kappa_{3 \beta} \hat{Q}_{3L} \hat{\rho}^{\prime} \hat{d}^{\prime c}_{ \beta L}+
\kappa^{\prime}_{5 \alpha} \hat{Q}_{\alpha L} \hat{\rho} \hat{u}^{\prime c}_{L} \nonumber \\ &+&
\kappa^{\prime}_{6 \alpha \beta} \hat{Q}_{\alpha L} \hat{\eta} \hat{d}^{\prime c}_{ \beta L}+
\kappa^{\prime}_{7 \alpha \beta} \hat{Q}_{\alpha L} \hat{\phi} \hat{d}^{\prime c}_{ \beta L}+ 
\kappa_{9 \alpha ai} \hat{Q}_{\alpha L} \hat{L}_{aL} \hat{d}^{c}_{iL} 
\nonumber \\ &+& 
\kappa^{\prime}_{9a \alpha \beta}\hat{L}_{aL} \hat{Q}_{\alpha L} \hat{d}^{\prime c}_{\beta L}+
\xi_{1ijk} \hat{d}^{c}_{iL} \hat{d}^{c}_{jL} \hat{u}^{c}_{kL}+
\xi_{2ij} \hat{d}^{c}_{iL} \hat{d}^{c}_{jL} \hat{u}^{\prime c}_{L} \nonumber \\ &+&
\xi_{3i \beta j} \hat{d}^{c}_{iL} \hat{d}^{\prime c}_{\beta L} \hat{u}^{c}_{jL}+
\xi_{4i \beta j} \hat{d}^{c}_{iL} \hat{d}^{\prime c}_{\beta L} \hat{u}^{\prime c}_{L}+
\xi_{5 \alpha \beta i} \hat{d}^{\prime c}_{\alpha L}\hat{d}^{\prime c}_{\beta L} \hat{u}^{c}_{iL} \nonumber \\ &+&
\xi_{6 \alpha \beta} \hat{d}^{\prime c}_{\alpha L}\hat{d}^{\prime c}_{\beta L}
\hat{u}^{\prime c}_{L}+
\xi_{7ij \beta} \hat{u}^{c}_{iL} \hat{u}^{c}_{jL} \hat{j}^{c}_{\beta L}+
\xi_{8i \beta} \hat{u}^{c}_{iL}\hat{u}^{\prime c}_{L}\hat{j}^{c}_{\beta L} 
\nonumber \\ &+&
\xi_{9 \beta} \hat{u}^{\prime c}_{L}\hat{u}^{\prime c}_{L}\hat{j}^{c}_{\beta L}+
\xi_{10i \beta} \hat{d}^{c}_{iL} \hat{J}^{c}_{L} \hat{j}^{c}_{\beta L}. 
\label{sprv}
\end{eqnarray}

The pattern of the symmetry breaking in this  model is given by

\begin{eqnarray}
&\mbox{susy341}& \stackrel{{\cal L}_{soft}}{\longmapsto}
\mbox{SU(3)}_C\ \times \ \mbox{SU(4)}_L\times \mbox{U(1)}_N
\stackrel{\langle\chi\rangle \langle
\chi^{\prime}\rangle}{\longmapsto} \mbox{SU(3)}_C \ \times \
\mbox{SU(2)}_L\times
\mbox{U(1)}_Y \nonumber \\
&\stackrel{\langle\rho,\eta,\phi,H \rho^{\prime}\eta^{\prime},\phi^{\prime},H^{\prime}\rangle}{\longmapsto}&
\mbox{SU(3)}_C \ \times \ \mbox{U(1)}_Q.
\end{eqnarray}

The covariant derivative for quadriplet ($4$) is defined as
\begin{equation}
{\cal D}^{m}\phi_{i}= \partial^{m}\phi_{i}+ \imath g \left( \vec{V}^{m} \cdot \frac{\vec{\lambda}}{2} 
\right)^{j}_{i}\phi_{j}+ \imath g^{\prime}N_{\phi}V_{m}\phi_{i},
\label{covder4} 
\end{equation}
and for the anti-quadriplet ($4^{*}$) we have
\begin{equation}
\overline{{\cal D}}^{m}\phi_{i}= \partial^{m}\phi_{i}+ \imath g \left( \vec{V}^{m} \cdot \frac{\vec{\bar{\lambda}}}{2} 
\right)^{j}_{i}\phi_{j}+ \imath g^{\prime}N_{\phi}V_{m}\phi_{i},
\label{covder4*} 
\end{equation}
where $g$ and $g^{\prime}$ are the $SU(4)_{L}$ and $U(1)_{N}$ gauge coupling constant respectivelly.

\section{Fermion Masses}

The terms in our Superpotencial necessary for give mass for the leptons are \cite{susy341,Rodriguez:2007jc}
\begin{eqnarray}
W^{lept}_{3}&=& \lambda_{2ab} \epsilon \hat{L}_{aL} \hat{L}_{bL} \hat{ \eta}+
\lambda_{3ab} \epsilon \hat{L}_{aL} \hat{L}_{bL} \hat{ \phi}.  \nonumber \\
\label{sp3l}
\end{eqnarray}
Those terms give the following Yukawa couplings
\begin{eqnarray}
{\cal L}^{lept}_{3}&=&- \left[ \lambda_{2ab} \epsilon L_{aL} L_{bL} \eta +
\lambda_{3ab} \epsilon L_{aL} L_{bL} \phi + 
\lambda_{4ab} L_{aL} L_{bL} H \right].  \nonumber \\
\label{yukawatermsleptonsmsusy341}
\end{eqnarray}
The last term is the same as presented in our Eq.(\ref{m4eq4358}), therefore our leptons get their masses and they are Dirac fermions in similar way 
as presented in the non supersymmetric 341 model.

The source for the mass for the quarks are \cite{susy341,Rodriguez:2007jc}
\begin{eqnarray}
W^{quarks}_{3}&=& \kappa_{1i} \hat{Q}_{3L} \hat{\eta}^{\prime} \hat{u}^{c}_{iL}+
\kappa^{\prime}_{1} \hat{Q}_{3L} \hat{\eta}^{\prime} \hat{u}^{\prime c}_{L}+
\kappa_{2i} \hat{Q}_{3L} \hat{\phi}^{\prime} \hat{u}^{c}_{iL} +
\kappa^{\prime}_{2}\hat{Q}_{3L} \hat{\phi}^{\prime} \hat{u}^{\prime c}_{L} +
\kappa_{3i} \hat{Q}_{3L} \hat{\rho}^{\prime} \hat{d}^{c}_{iL} \nonumber \\ &+&
\kappa_{3 \beta} \hat{Q}_{3L} \hat{\rho}^{\prime} \hat{d}^{\prime c}_{ \beta L}+
\kappa_{4} \hat{Q}_{3L} \hat{\chi}^{\prime} \hat{J}^{c}_{L}+
\kappa_{5 \alpha i} \hat{Q}_{\alpha L} \hat{\rho} \hat{u}^{c}_{iL} +
\kappa^{\prime}_{5 \alpha} \hat{Q}_{\alpha L} \hat{\rho} \hat{u}^{\prime c}_{L} +
\kappa_{6 \alpha i} \hat{Q}_{\alpha L} \hat{\eta} \hat{d}^{c}_{iL} \nonumber \\ &+&
\kappa^{\prime}_{6 \alpha \beta} \hat{Q}_{\alpha L} \hat{\eta} \hat{d}^{\prime c}_{ \beta L}+
\kappa_{7 \alpha i} \hat{Q}_{\alpha L} \hat{\phi} \hat{d}^{c}_{iL} +
\kappa^{\prime}_{7 \alpha \beta} \hat{Q}_{\alpha L} \hat{\phi} \hat{d}^{\prime c}_{ \beta L}+ 
\kappa_{8 \alpha \beta} \hat{Q}_{\alpha L} \hat{\chi} \hat{j}^{c}_{\beta L},  \nonumber \\
\label{sp3q}
\end{eqnarray}
they generate the following Yukawa terms
\begin{eqnarray}
{\cal L}^{quarks}_{Yukawa}&=&- \left[ \kappa_{1i} \left( Q_{3L} \eta^{\prime} \right) u^{c}_{iL}+
\kappa^{\prime}_{2} \left( Q_{3L} \phi^{\prime} \right)  u^{\prime c}_{L} +
\kappa_{4}\left(  Q_{3L} \chi^{\prime} \right) J^{c}_{L}+
\kappa^{\prime}_{5 \alpha} \left( Q_{\alpha L} \rho \right) u^{\prime c}_{L} \right. \nonumber \\ 
&+& \left.
\kappa_{6 \alpha i} \left( Q_{\alpha L} \eta \right) d^{c}_{iL} +
\kappa^{\prime}_{6 \alpha \beta} \left( Q_{\alpha L} \eta \right) d^{\prime c}_{ \beta L}+
\kappa^{\prime}_{7 \alpha \beta} \left( Q_{\alpha L} \phi \right) d^{\prime c}_{ \beta L} \right],
\end{eqnarray}
compare it with Eq.(\ref{m4eq468}).

\section{Boson Masses}

We get gauge bosons masses in the followin term
\begin{eqnarray}
&&\left( {\cal D} \langle\eta\rangle \right)^{\dagger}\left( {\cal D} \langle\eta\rangle \right)+
\left( {\cal D} \langle \rho\rangle \right)^{\dagger}\left( {\cal D} \langle \rho\rangle \right)+
\left( {\cal D} \langle \phi \rangle \right)^{\dagger}\left( {\cal D} \langle \phi \rangle \right)+
\left( {\cal D} \langle \chi \rangle \right)^{\dagger}\left( {\cal D} \langle \rho\rangle \right)+
\left( {\cal D}\langle H \rangle \right)^{\dagger}\left( {\cal D} \langle H\rangle \right) \nonumber \\ &+&
\left( {\cal D} \langle\eta^{\prime}\rangle \right)^{\dagger}\left( {\cal D} \langle\eta^{\prime}\rangle \right)+
\left( {\cal D} \langle \rho^{\prime}\rangle \right)^{\dagger}\left( {\cal D} \langle \rho^{\prime}\rangle \right)+
\left( {\cal D} \langle \phi^{\prime} \rangle \right)^{\dagger}\left( {\cal D} \langle \phi^{\prime}\rangle \right)+
\left( {\cal D} \langle \chi^{\prime} \rangle \right)^{\dagger}\left( {\cal D} \langle \chi^{\prime}\rangle \right) \nonumber \\ &+&
\left( {\cal D}\langle H^{\prime} \rangle \right)^{\dagger}\left( {\cal D} \langle  H^{\prime}\rangle \right)
\end{eqnarray}

Using Eq.(\ref{gaugebosonsu(4)quad}), we can write
\begin{eqnarray}
{\cal D}^{m} \langle \eta \rangle &=& \frac{- \imath gv}{\sqrt{2}}\left(
\begin{array}{c}
D_{1}^{m} \\
W^{- m} \\
X^{0 m} \\
V^{+}_{2 m}
\end{array}
\right), \,\
{\cal D}^{m} \langle \phi \rangle = \frac{- \imath gz}{\sqrt{2}}\left(
\begin{array}{c}
\left( X^{0 m} \right)^{*} \\
V^{-}_{1 m} \\
D_{3}^{m} \\
V^{+}_{3 m}
\end{array}
\right), \nonumber \\
\end{eqnarray}
\begin{eqnarray}
{\cal D}^{m} \langle \rho \rangle &=& \frac{- \imath gu}{\sqrt{2}}\left(
\begin{array}{c}
W^{+ m} \\
D_{2}^{m} \\
V^{+}_{1 m} \\
U^{++}_{ m}
\end{array}
\right) - \imath g^{\prime}uV^{m}, \nonumber \\
\overline{{\cal D}}^{m} \langle \chi \rangle &=& \frac{- \imath gw}{\sqrt{2}}\left(
\begin{array}{c}
V^{-}_{2 m} \\
U^{--}_{ m} \\
V^{-}_{3 m}  \\
D_{4}^{m}  
\end{array}
\right) - \imath g^{\prime}\left( -1 \right)wV^{m}, \nonumber \\. \nonumber  
\label{dc4plet} 
\end{eqnarray}

Using Eq.(\ref{gaugebosonsu(4)antiquad}), we get
\begin{eqnarray}
{\cal D}^{m} \langle \eta^{\prime} \rangle &=& \frac{- \imath gv^{\prime}}{\sqrt{2}}\left(
\begin{array}{c}
D_{1}^{m} \\
W^{+ m} \\
\left( X^{0 m} \right)^{*} \\
V^{-}_{2 m}
\end{array}
\right), \,\
{\cal D}^{m} \langle \phi^{\prime} \rangle = \frac{- \imath gz^{\prime}}{\sqrt{2}}\left(
\begin{array}{c}
X^{0 m}  \\
V^{+}_{1 m} \\
D_{3}^{m} \\
V^{-}_{3 m}
\end{array}
\right), \nonumber \\
\end{eqnarray}
\begin{eqnarray}
{\cal D}^{m} \langle \rho^{\prime} \rangle &=& \frac{- \imath gu^{\prime}}{\sqrt{2}}\left(
\begin{array}{c}
W^{- m} \\
D_{2}^{m} \\
V^{-}_{1 m} \\
U^{--}_{ m}
\end{array}
\right) - \imath g^{\prime} \left( -1 \right) u^{\prime}V^{m}, \nonumber \\
\overline{{\cal D}}^{m} \langle \chi^{\prime} \rangle &=& \frac{- \imath gw^{\prime}}{\sqrt{2}}\left(
\begin{array}{c}
V^{+}_{2 m} \\
U^{++}_{ m} \\
V^{+}_{3 m}  \\
D_{4}^{m}  
\end{array}
\right) - \imath g^{\prime}w^{\prime}V^{m}, \nonumber \\. \nonumber  
\label{dca4plet} 
\end{eqnarray}

We can show the masses of the gauge bosons, see Eqs.(\ref{dc4plet},\ref{dca4plet}), are given by
The electroweak gauge bosons of this theory are defined at Eq.(\ref{defcharggaugebosons}) and they get the followinghave masses 
\begin{eqnarray}
M^{2}_{W}&=&\frac{g^{2}}{4} \left[
\left( v^{2}+ v^{\prime 2} \right)+ \left( u^{2}+ u^{\prime 2} \right)+2 \left( x^{2}+ x^{\prime 2} \right) 
\right], \nonumber \\
M^{2}_{V_{1}}&=&\frac{g^{2}}{4} \left[
\left( z^{2}+ z^{\prime 2} \right)+ \left( u^{2}+ u^{\prime 2} \right)+2 \left( x^{2}+ x^{\prime 2} \right)
\right] , \nonumber \\
M^{2}_{V_{2}}&=&\frac{g^{2}}{4} \left[
\left( v^{2}+ v^{\prime 2} \right)+ \left( w^{2}+ w^{\prime 2} \right)+2 \left( x^{2}+ x^{\prime 2} \right)
\right], \nonumber \\
M^{2}_{V_{3}}&=&\frac{g^{2}}{4}\left[
\left( z^{2}+ z^{\prime 2} \right)+ \left( w^{2}+ w^{\prime 2} \right)+2 \left( x^{2}+ x^{\prime 2} \right)
\right], \nonumber \\
M^{2}_{X}&=& \frac{g^{2}}{8}\left[
\left( v^{2}+ v^{\prime 2} \right)+ \left( z^{2}+ z^{\prime 2} \right)
\right],  \nonumber \\
M^{2}_{U}&=& \frac{g^{2}}{4}\left[
\left( v^{2}+ v^{\prime 2} \right)+ \left( w^{2}+ w^{\prime 2} \right)+4 \left( z^{2}+ z^{\prime 2} \right)
\right].
\end{eqnarray}
In the case of the neutral gauge bosons we get a similar mass matrix showed in 
our Eq.(\ref{zmass}).

\section{Conclusions}
\label{sec:con}

We have built the complete supersymmetric version of the 341 model of Ref.
\cite{su4b,su4b1,su4b2}. We have also obtained the mass spectrum for the Gauge 
Bosons and also for the usual fermions in this model.

\begin{center}
{\Large {\bf Acknowledgments}}
\end{center} 
This work was supported by Conselho Nacional de Ci\^encia e Tecnologia (CNPq) 
under the processes 309564/2006-9.

\appendix

\section{The generators of $SU(4)$}
\label{sec:gensu(4)}

The Hermitian generators of $SU(4)$ are given by \cite{greiner}:
\begin{eqnarray}
\lambda_{1}&=& \left( 
\begin{array}{cccc}
0 & 1 & 0 & 0 \\
1 & 0 & 0 & 0 \\
0 & 0 & 0 & 0 \\
0 & 0 & 0 & 0
\end{array}
\right) \,\
\lambda_{2}= \left( 
\begin{array}{cccc}
0 &- \imath & 0 & 0 \\
\imath & 0 & 0 & 0 \\
0 & 0 & 0 & 0 \\
0 & 0 & 0 & 0
\end{array}
\right) \,\
\lambda_{3}= \left( 
\begin{array}{cccc}
1 & 0 & 0 & 0 \\
0 & -1 & 0 & 0 \\
0 & 0 & 0 & 0 \\
0 & 0 & 0 & 0
\end{array}
\right) \nonumber \\
\lambda_{4}&=& \left( 
\begin{array}{cccc}
0 & 0 & 1 & 0 \\
0 & 0 & 0 & 0 \\
1 & 0 & 0 & 0 \\
0 & 0 & 0 & 0
\end{array}
\right) \,\
\lambda_{5}= \left( 
\begin{array}{cccc}
0 & 0 &- \imath & 0 \\
0 & 0 & 0 & 0 \\
\imath & 0 & 0 & 0 \\
0 & 0 & 0 & 0
\end{array}
\right) \,\
\lambda_{6}= \left( 
\begin{array}{cccc}
0 & 0 & 0 & 0 \\
0 & 0 & 1 & 0 \\
0 & 1 & 0 & 0 \\
0 & 0 & 0 & 0
\end{array}
\right) \nonumber \\
\lambda_{7}&=& \left( 
\begin{array}{cccc}
0 & 0 & 0 & 0 \\
0 & 0 &- \imath & 0 \\
0 & \imath & 0 & 0 \\
0 & 0 & 0 & 0
\end{array}
\right) \,\
\lambda_{8}= \frac{1}{\sqrt{3}} \left( 
\begin{array}{cccc}
1 & 0 & 0 & 0 \\
0 & 1 & 0 & 0 \\
0 & 0 &- 2 & 0 \\
0 & 0 & 0 & 0
\end{array}
\right) \,\
\lambda_{9}= \left( 
\begin{array}{cccc}
0 & 0 & 0 & 1 \\
0 & 0 & 1 & 0 \\
0 & 1 & 0 & 0 \\
1 & 0 & 0 & 0
\end{array}
\right) \nonumber \\
\lambda_{10}&=& \left( 
\begin{array}{cccc}
0 & 0 & 0 &- \imath \\
0 & 0 & 1 & 0 \\
0 & 1 & 0 & 0 \\
\imath & 0 & 0 & 0
\end{array}
\right) \,\
\lambda_{11}= \left( 
\begin{array}{cccc}
0 & 0 & 0 & 0 \\
0 & 0 & 0 & 1 \\
0 & 0 & 0 & 0 \\
0 & 1 & 0 & 0
\end{array}
\right) \,\
\lambda_{12}= \left( 
\begin{array}{cccc}
0 & 0 & 0 & 0 \\
0 & 0 & 0 &- \imath \\
0 & 0 & 0 & 0 \\
0 & \imath & 0 & 0
\end{array}
\right) \nonumber \\
\lambda_{13}&=& \left( 
\begin{array}{cccc}
0 & 0 & 0 & 0 \\
0 & 0 & 1 & 0 \\
0 & 1 & 0 & 1 \\
0 & 0 & 1 & 0
\end{array}
\right) \,\
\lambda_{14}= \left( 
\begin{array}{cccc}
0 & 0 & 0 & 0 \\
0 & 0 & 1 & 0 \\
0 & 1 & 0 &- \imath \\
0 & 0 & \imath & 0
\end{array}
\right) \,\
\lambda_{15}= \frac{1}{\sqrt{6}} \left( 
\begin{array}{cccc}
1 & 0 & 0 & 0 \\
0 & 1 & 0 & 0 \\
0 & 0 & 1 & 0 \\
0 & 0 & 0 &- 3
\end{array}
\right) \nonumber \\ 
\label{geradoresdesu(4)}
\end{eqnarray}

Using the matrices above we can write
\begin{eqnarray}
\sum_{i=1}^{15}V_{i}^{m} \left( \frac{\lambda_{i}}{2} \right) = \frac{1}{\sqrt{2}} \left( 
\begin{array}{cccc}
D_{1}^{m} & W^{+ m} & X^{0 m} & V^{- m}_{2} \\
W^{- m} & D_{2}^{m} & V^{- m}_{1} & U^{-- m} \\
\left(  X^{0 m} \right)^{*} & V^{+ m}_{1} & D_{3}^{m} & V^{- m}_{3} \\
V^{+ m}_{2} & U^{++ m} & V^{+ m}_{3} & D_{4}^{m}
\end{array}
\right), \nonumber \\ 
\label{gaugebosonsu(4)quad}
\end{eqnarray}
where we have defined
\begin{eqnarray}
W^{\pm}&=& \frac{1}{\sqrt{2}} \left( V^{1} \mp \imath V^{2} \right), \,\
V^{\pm}_{1}= \frac{1}{\sqrt{2}} \left( V^{6} \pm \imath V^{7} \right), \nonumber \\
V^{\pm}_{2}&=& \frac{1}{\sqrt{2}} \left( V^{9} \pm \imath V^{10} \right), \,\
V^{\pm}_{3}= \frac{1}{\sqrt{2}} \left( V^{13} \pm \imath V^{14} \right), 
\nonumber \\
U^{\pm \pm}&=& \frac{1}{\sqrt{2}} \left( V^{11} \pm \imath V^{12} \right), \,\
X^{0}= \frac{1}{\sqrt{2}} \left( V^4+ \imath V^{5} \right),
\label{defcharggaugebosons}
\end{eqnarray}
is the $W$-boson in the SM as is usual and we also defined
we have defined
\begin{eqnarray}
D_{1}^{m}&=& V^{m}_{3}+ \frac{V^{m}_{8}}{\sqrt{3}}+ \frac{V^{m}_{15}}{\sqrt{6}},
\nonumber \\
D_{2}^{m}&=&- V^{m}_{3}+ \frac{V^{m}_{8}}{\sqrt{3}}+ \frac{V^{m}_{15}}{\sqrt{6}},
\nonumber \\
D_{3}^{m}&=&- \frac{2V^{m}_{8}}{\sqrt{3}}+ \frac{V^{m}_{15}}{\sqrt{6}},
\nonumber \\
D_{4}^{m}&=&- \frac{3V^{m}_{15}}{\sqrt{6}}.
\label{basefotonz-zp-zpp}
\end{eqnarray}

We can also write
\begin{eqnarray}
\sum_{i=1}^{15}V_{i}^{m} \left( \frac{\lambda_{i*}}{2} \right) = \frac{1}{\sqrt{2}} \left( 
\begin{array}{cccc}
D_{1}^{m} & W^{- m} & \left(  X^{0 m} \right)^{*} & V^{+ m}_{2} \\
W^{+ m} & D_{2}^{m} & V^{+ m}_{1} & U^{++ m} \\
X^{0 m} & V^{- m}_{1} & D_{3}^{m} & V^{+ m}_{3} \\
V^{- m}_{2} & U^{-- m} & V^{- m}_{3} & D_{4}^{m}
\end{array}
\right), \nonumber \\ 
\label{gaugebosonsu(4)antiquad}
\end{eqnarray}

\end{document}